\documentstyle[11pt,newpasp,twoside,epsf]{article}

\begin{document}
\title{Predictions on the number of variable stars for
            the GAIA space mission
      and for surveys as the ground-based
      International Liquid Mirror Telescope}
\author{Laurent Eyer}
\affil{Instituut voor Sterrenkunde, K.U.Leuven,
B-3001 Heverlee, Belgium}
\author{Jan Cuypers}
\affil{Royal Observatory of Belgium, Ringlaan~3,
B-1180 Brussel, Belgium}

\begin{abstract}
Future space and ground-based survey programmes will produce an impressive
amount of photometric data. The GAIA space mission will map the
complete sky down to mag V=20 and produce time series for about 1
billion stars. Survey instruments as the International Liquid Mirror Telescope will
observe slices of the sky down to magnitude V=23.
In both experiments, the opportunity exists to discover a huge amount
of variable stars. A prediction of
the expected total number of variable stars and the number of
variables in specific subgroups is given.
\end{abstract}
\vspace{-0.7truecm}
\section{The total number of variable stars} A first estimate of the
total number of variable stars observable by GAIA was done by Eyer
(1999). The star population used came from the star-count model of
Figueras~et~al.~(1999) and the variability detection threshold was
derived from the Hipparcos survey results. With the new
qualifications of the GAIA mission, about 1 billion stars (up to
mag~G$<$20) are expected to be observed, with about 18~million
variable stars, including about 5~million "classic" periodic
variables.\\
Very different star counts are obtained according to the extinction
laws used (Figueras, private communication). Since the quality of the
GAIA photometry in the crowded fields is still uncertain, we cannot
discuss here the number of variables in dense clusters and
galaxies.\\
About 2 to 3 millions eclipsing binaries will be observed, but their
detection probability will be studied in detail in the future. About
300\,000 stars with rotation induced variability can be expected as
well.
\section{The methods} For a specific interval of V-I, we computed the
proportion of variables in the Hipparcos survey and we applied that
rate to the number of stars obtained from the Figueras model
(method~A). Surface densities were calculated, either from the
Hipparcos parallaxes or from the specific properties of the stars. We
integrated and removed the stars behind the bulge (method~B). We
extrapolated the GCVS data (Kholopov et al., 1998) assuming detection
completeness up to a certain magnitude and a magnitude limit for the
population beyond which no more stars are present (method~C). We also
analysed the detection rates of the microlensing surveys (when
available) and scanned the literature.
\section{Pulsating variables} Method
A and C estimate the number of $\beta$ Cephei stars to be about 3000.
15\,000 SPB variables will be detected according to Method A.
Applying method A and C gave about the same estimate for $\delta$
Scuti stars: 60\,000. However, it will be very difficult to analyse
the very reddened low amplitude variables. With
method B even higher numbers as 240\,000 $\delta$~Scuti stars show up.\\
With a total number of RR Lyrae as given by  Suntzeff et al. (1991)
we arrive at 70\,000 observable RR~Lyrae (method~B). Using the OGLE
and MACHO detection rates, we expect 15\,000 to 40\,000 RR Lyrae in
the bulge.\\
All galactic Cepheids are within the observational range of GAIA, if
not too obscured by interstellar extinction. Results of recent deep
surveys confirm the early estimates of a total of 2000 to 8000
Cepheids. With the help of the Fernie database (1995), we
obtained (Method B) a density of 15-20
Cepheids/$\mbox{kpc}^2$, leading to an estimate of 5\,200--6\,900
observable stars.\\
Early estimates gave in total 200\,000 Mira and related long period
variables in the Galaxy. With
500~Miras/$\mbox{kpc}^2$, 140\,000 to 170\,000 Miras will be
observable. Method B gave us a density of 250-350 Semi-Regular
variables/$\mbox{kpc}^2$ or a total of 100\,000 observable SR
stars.\\
We plan to calculate and analyse all categories of variables
stars in more detail to arrive at reliable estimates of all
observable variable stars in the Galaxy.
\section{Variable stars in deep surveys} An example:
The International Liquid MirrorTelescope (ILMT).\\
(see \verb=http://vela.astro.ulg.ac.be/themes/telins/lmt/index_e.html=) \\
An international group of institutions  are actively interested in
developing a 4-m class liquid mirror telescope.  If the view of the
ILMT includes fields near the galactic center and all stars from R
magnitude 17 up to 20 can be measured with high precision ($\leq$ 0.01
mag), the project will yield a unique time series of about 2 million
stars with a total of 500 measurements of each star during 5 years.
About 10\,000 new variable stars can be expected, including 6\,000 faint
eclipsing binaries, 200 RR Lyrae and 300 long periodic variables.
\nopagebreak


\begin{references}
\reference Eyer, L. 1999, Balt. Ast., 8, 321
\reference Fernie, J.D., Beattie, B., Evans, N.R., \& Seager, S. 1995, IBVS 4148
\reference Figueras, F. et al., 1999, Balt. Ast., 8, 291
\reference GAIA: \verb=http://astro.estec.esa.nl/SA-general/Projects/GAIA/=
\reference HIPPARCOS: Hipparcos and Tycho catalogues, ESA SP-1200
\reference Kholopov, P.N., et al. 1998, GCVS, 4th Edition
\reference Suntzeff, N.B., Kinman, T.D., \& Kraft, R.P. 1991, \apj, 367, 528
\end{references}
\end{document}